\documentclass[11pt,oneside]{article}
\usepackage{multicol}
\usepackage{PRIMEarxiv}

\usepackage[utf8]{inputenc} 
\usepackage[T1]{fontenc}    
\usepackage{hyperref}       
\usepackage{url}            
\usepackage{booktabs}       
\usepackage{amsfonts}  
\usepackage{amsmath}        
\usepackage{setspace}
\usepackage{nicefrac}       
\usepackage{microtype}      
\usepackage{lipsum}
\usepackage{fancyhdr}
\usepackage{graphicx}       
\graphicspath{{media/}}     

\usepackage{authblk}

\title{Assessing Reusability of Deep Learning-Based Monotherapy Drug Response Prediction Models Trained with Omics Data}

\author[1]{Jamie C. Overbeek$^*$}
\author[1]{Alexander Partin$^*$} 
\author[1]{Thomas S. Brettin}
\author[1]{Nicholas Chia}
\author[1]{Oleksandr Narykov}
\author[1]{Priyanka Vasanthakumari}
\author[1]{Andreas Wilke}
\author[1]{Yitan Zhu}
\author[3]{Austin Clyde}
\author[4]{Sara Jones}
\author[5]{Rohan Gnanaolivu}
\author[5]{Yuanhang Liu}
\author[5]{Jun Jiang}
\author[5]{Chen Wang}
\author[6]{Carter Knutson}
\author[6]{Andrew McNaughton}
\author[6]{Neeraj Kumar}
\author[7]{Gayara Demini Fernando}
\author[7]{Souparno Ghosh}
\author[8]{Cesar Sanchez-Villalobos}
\author[8]{Ruibo Zhang}
\author[8]{Ranadip Pal}
\author[4]{M. Ryan Weil}
\author[2,3]{Rick L. Stevens}

\affil[1]{Data Science and Learning Division, Argonne National Laboratory, Lemont, IL, USA}
\affil[2]{Computing, Environment and Life Sciences, Argonne National Laboratory, Lemont, IL, USA}
\affil[3]{Department of Computer Science, The University of Chicago, Chicago, IL, USA}
\affil[4]{Frederick National Laboratory for Cancer Research, Frederick, MD, USA}
\affil[5]{Department of Quantitative Health Sciences, Mayo Clinic, Rochester, MN, USA}
\affil[6]{Pacific Northwest National Laboratory, Richland, WA, USA}
\affil[7]{Department of Statistics, University of Nebraska–Lincoln, Lincoln, NE, USA}
\affil[8]{Department of Electrical \& Computer Engineering, Texas Tech University, Lubbock, TX, USA}

\begin{document}
\maketitle
\def\thefootnote{*}
\footnotetext{These authors contributed equally to this work. Please cite as Overbeek JC and Partin A et al. Correspondence: joverbeek@anl.gov, apartin@anl.gov}

\begin{abstract}


%
Cancer drug response prediction (DRP) models present a promising approach towards precision oncology, tailoring treatments to individual patient profiles.
While deep learning (DL) methods have shown great potential in this area, models that can be successfully translated into clinical practice and shed light on the molecular mechanisms underlying treatment response will likely emerge from collaborative research efforts. 
This highlights the need for reusable and adaptable models that can be improved and tested by the wider scientific community.
%
In this study, we present a scoring system for assessing the reusability of prediction DRP models, and apply it to 17 peer-reviewed DL-based DRP models. As part of the IMPROVE (Innovative Methodologies and New Data for Predictive Oncology Model Evaluation) project, which aims to develop methods for systematic evaluation and comparison DL models across scientific domains, we analyzed these 17 DRP models focusing on three key categories: software environment, code modularity, and data availability and preprocessing. While not the primary focus, we also attempted to reproduce key performance metrics to verify model behavior and adaptability.
%
Our assessment of 17 DRP models reveals both strengths and shortcomings in model reusability. Issues like underspecified software environments, lack of workflow documentation, and omission of data preprocessing scripts create challenges for users attempting to reproduce or build upon these models. 
To promote rigorous practices and open-source sharing, we offer recommendations for developing and sharing prediction models. Following these recommendations can address many of the issues identified in this study, improving model reusability without adding significant burdens on researchers.
%
This work offers the first comprehensive assessment of reusability and reproducibility across diverse DRP models, providing insights into current model sharing practices and promoting standards within the DRP and broader AI-enabled scientific research community. The community will develop further guidelines as engagement with this topic increases.
\end{abstract}


\section{Introduction} \label{sec:intro}

There is a growing interest in utilizing predictive modeling to advance precision oncology \cite{senft_precision_2017, ballester_artificial_2021}. 
Among various approaches, drug response prediction (DRP) models hold promise, as they predict the response of cancer to drug treatments by analyzing tumor and drug properties \cite{ballester_artificial_2021-1}.
Technological advances in genomic profiling and drug sensitivity screenings have led to an abundance of data \cite{piyawajanusorn_gentle_2021, chiu_deep_2020}, driving the exploration of machine learning (ML) and especially deep learning (DL) methods, for DRP \cite{partin_deep_2023, firoozbakht_overview_2022, adam_machine_2020}.
Numerous papers have been published, exploring various model architectures, training schemes, and data representations, demonstrating the predictive performance of different approaches in retrospective analyses with preclinical drug screening data \cite{partin_deep_2023}. 
While this surge of publications represents great progress in the field, it also makes it extremely challenging to navigate the various methodologies and identify the most effective approaches to improve the utility of DRP models in precision oncology.

To demonstrate superiority, most existing papers compare the performance of their DL models against conventional, out-of-the-box ML baselines \cite{partin_deep_2023}. 
However, to demonstrate substantive contributions, newly proposed models should undergo rigorous cross-comparisons with multiple state-of-the-art models from the research community.
Such comparisons typically require models to use consistent data representations, data splits, and evaluation schemes, which often necessitate modifications to the original code of the models. 
Therefore, the ability to efficiently integrate community models into diverse comparison analyses is directly linked to model \textit{reusability}. 
While several studies have performed comprehensive model comparisons focusing on prediction performance \cite{costello_community_2014, menden_community_2019, chen_survey_2021}, there is minimal discussion provided about the reusability of these models.

Given the potential challenges of working with community models, having insights into the ease of using and \textit{reusing} models can substantially influence researchers' decision-making process when selecting models for comparison studies or adapting them to novel contexts.
So et al. \cite{so2023reusability} conducted extensive analysis with TCRP \cite{ma2021tcrp}, assessing its reusability on 'omics and drug response datasets beyond the original publication, including cell line \cite{safikhani2017gene, thu2018disruption}, patient-derived xenograft \cite{gao2015high}, and clinical trial data \cite{hatzis2011genomic, itoh2014estrogen, baldasici2022circulating, horak2013biomarker}.
Their conclusion of TCRP's successful reusability primarily relied on its prediction performance across contexts.
However, the challenges encountered during model adaptation can play an equally critical role in model selection.
While providing extensive discussion on these challenges, a concrete system of criteria assessing ease of use was lacking.
A set of well-defined criteria for assessing reusability can facilitate efficient model selection.
These criteria could take on a form of reusability ranking, evaluating models based on factors relating to code accessibility and modularity, quality of documentation, and compatibility with popular frameworks.
This would allow researchers to quickly identify models that are not only high-performing but also user-friendly, fostering greater collaboration and innovation in AI-assisted precision oncology.

In this paper, we propose a scoring system to assess model reusability and apply it to 17 DRP models from peer-reviewed publications.
This work is part of the IMPROVE project (Innovative Methodologies and New Data for Predictive Oncology Model Evaluation)\footnote{https://computational.cancer.gov/about/improve}, which aims to develop computational methods for the systematic evaluation and comparison of DL models across scientific domains, with DRP as a driving use case.
From over a hundred published DL-based DRP models \cite{partin_deep_2023}, we selected 17 based on the year of publication, deep learning framework used, and other criteria outlined in detail in Section \ref{sec:selection_criteria}.
Each researcher from our collaborative team analyzed one or two models.
The reusability assessment consisted of three major categories:
\begin{enumerate}
    \item \textbf{Software Environment} (Section \ref{sec:sw_env_methods}): Scoring models based on the availability of information for setting up the computational environment.
    \item \textbf{Code Modularity} (Section \ref{sec:modularity_methods}): Evaluating the ability to execute major ML steps individually, including data preprocessing, model training, and inference, and assessing the description of this modularity in the documentation.
    \item \textbf{Data Availability and Preprocessing} (Section \ref{sec:data_and_preproc_methods}): Assessing the availability of data and preprocessing code that transforms raw data into model-ready files.
\end{enumerate}
We aimed to reproduce key performance metrics for each model to ensure expected behavior and adaptability for new contexts (Section \ref{sec:reproduce_methods}).

This study represents the first comprehensive assessment of model reusability and reproducibility across diverse DRP models. Evaluating 17 peer-reviewed models provides insights into current model sharing and documentation practices in the DRP research community. Our findings aim to highlight model sharing challenges and promote standards for sharing DRP and AI-enabled scientific models.

\begin{table}[t]
\centering
\resizebox{\columnwidth}{!}{
\begin{tabular}{@{}clllllll@{}}
\toprule
 & Model & Year & Framework & Cancer 'Omics Features & Drug Features & Treatment Response \\
\midrule
1 & BiG-DRP \cite{hostallero2022big-drp} & 2022 & PyTorch & GE & DD & IC50 \\ 
2 & DeepCDR \cite{liu2020deepcdr} & 2020 & TF-Keras & GE, Methyl, Mu & MG & IC50, Bin-IC50 \\ 
3 & DeepTTA \cite{jiang2022deeptta} & 2022 & PyTorch & GE & ESPF & IC50 \\
4 & DRPreter \cite{shin2022drpreter} & 2022 & PyTorch & GE & MG & IC50 \\ 
5 & DrugCell \cite{kuenzi2020drugcell} & 2020 & PyTorch & Mu & Morgan FP & AUC \\ 
6 & DrugGCN \cite{kim2021druggcn} & 2021 & TF & GE & None & IC50 \\
7 & DualGCN \cite{ma2022dualgcn} & 2022 & TF-Keras & CNV, GE & MG & IC50 \\ 
8 & GraphCDR \cite{liu2021graphcdr} & 2021 & PyTorch & GE, Methyl, Mu & MG & Bin-IC50 (GDSC), Bin-AAC (CCLE) \\
9 & GraphDRP \cite{nguyen2021graphdrp} & 2022 & PyTorch & CNV, Mu & MG & IC50 \\
10 & HiDRA \cite{jin2021hidra} & 2021 & TF-Keras & GE & Morgan FP & IC50 \\ 
11 & IGTD \cite{zhu2021igtd} & 2021 & TF-Keras & GE & DD & AUC \\
12 & MOLI \cite{sharifi2019moli} & 2019 & PyTorch & CNV, GE, Mu & None & Bin-IC50 (GDSC), Bin (PDXE, TCGA) \\ 
13 & PathDSP \cite{tang2021pathdsp} & 2021 & PyTorch & CNV, GE, Mu & Morgan FP & IC50 \\
14 & SWNet \cite{zuo2021swnet} & 2021 & PyTorch & GE, Mu & MG, Morgan FP & IC50 \\ 
15 & tCNNs \cite{liu2019tcnns} & 2019 & TF & CNV, Mu & SMILES (one-hot enc.) & IC50 \\
16 & TCRP \cite{ma2021tcrp} & 2021 & PyTorch & GE, Mu & None & AUC \\ 
17 & TGSA \cite{zhu2022tgsa} & 2022 & PyTorch & CNV, GE, Mu & MG & IC50 \\ 
\bottomrule
\end{tabular}
}
\caption{\textbf{Models and attributes used in this study.}
All models except DrugCell and TCRP were trained and predicted with the GDSC (Genomics of Drug Sensitivity in Cancer \cite{yang2012genomics}) dataset. DrugCell was trained on combined GDSC and CTRP (Cancer Therapeutic Response Portal \cite{seashore2015harnessing}) data, and TCRP was pre-trained on GDSC and fine-tuned on PDTC (patient-derived tumor cells).
\textbf{Cancer and Drug Features} include commonly used representations in drug response prediction models (note that some models incorporate unique representations not detailed in this table).
The \textbf{Treatment Response} indicates the outcome predicted by the model.
\textbf{Abbreviations}; TF, TensorFlow; GE, gene expression; Methyl, methylation; Mu, mutation; CNV, copy number variation; DD, drug descriptors; MG, molecular graph; ESPF, explainable substructure partition fingerprint; Morgan FP, Morgan fingerprint; SMILES, simplified molecular input line entry system; IC50, half-maximal inhibitory concentration; AUC, area under the dose-response curve; Bin, binary (i.e. response divided into sensitive and resistant); AAC, area above the dose-response curve.
}
\label{table:list_of_models}
\end{table}

\section{Methods} \label{sec:methods}

The Methods section outlines our approach for selecting community models and the subsequent analysis conducted on these models. Section \ref{sec:selection_criteria} details the criteria used for model selection, ensuring a representative and relevant subset. Sections \ref{sec:sw_env_methods} to \ref{sec:data_and_preproc_methods} describe the proposed reusability scoring system, which evaluates models based on software environment setup, code modularity, and data preprocessing. Additionally, we conducted a reproducibility analysis of key performance metrics to validate the models’ expected behavior and adaptability.

\subsection{Model selection criteria} \label{sec:selection_criteria}

We previously assembled a list of models that use DL methods for DRP, encompassing over one hundred original research papers. Given this collection, we utilize the following selection criteria to choose a subset of models for this study:

\begin{enumerate}
    \item The model is accompanied by open-source code, with preference to more recent publications (2019 onwards). 
    \item The model is deep learning-based and implemented in PyTorch or TensorFlow/Keras. 
    \item The model predicts monotherapy drug response values with end-to-end learning.
    \item The code repository associated with the model contains a basic description of the steps needed to use the model and reproduce the results.
\end{enumerate}

Following the aforementioned criteria, we selected a non-comprehensive set of 17 models that our team could successfully set up and run, summarized in Table \ref{table:list_of_models}. This is not an exhaustive list of all models meeting the selection criteria, but rather a subset feasible to analyze within our time constraints. Thus, the exclusion of any specific model was due to time limitations and not a judgment of its reusability.

\subsection{Software Environment} \label{sec:sw_env_methods}

Proper setup of the model's computational environment is crucial for reproducing the results as described in the original research papers. The environment setup typically involves installing required packages using Conda or pip, and often requires specific versions to avoid compatibility issues. The required software environment is generally detailed in a configuration file (e.g., requirements.txt, environment.yml) or described in a README file associated with the repository. Correctly setting up the environment ensures that the model functions as expected, while incomplete or outdated setups can lead to errors and multiple reinstallation attempts. Repositories that fail to provide complete environment information present a challenge, especially for users who may not have deep expertise in package management or troubleshooting.

Below is the scoring system used to assess the quality of the software environment information provided in the documentation of code repositories:

\textbf{Score of 4}.
The repository includes a requirements file or README that lists all necessary packages and specific version numbers. This detailed information allows users to set up the exact environment required, minimizing the risk of compatibility issues or unexpected behavior.

\textbf{Score of 3}.
The repository contains a requirements file or README with all necessary packages but with limited or no specification of version numbers. Although users can generally set up the environment, the lack of precise version details can lead to issues with incompatible package versions.

\textbf{Score of 2}.
The repository includes a requirements file or README with only a subset of the required packages. This incomplete list makes it difficult for users to set up the environment properly, potentially causing significant errors and requiring additional research to identify the missing packages.

\textbf{Score of 1}.
The repository contains no information on the required software environment. Users would have to infer or guess which packages are needed, likely resulting in significant reproducibility issues and extensive troubleshooting.

\subsection{Code Modularity} \label{sec:modularity_methods} 

Enhanced modularity in code repositories facilitates easier understanding of models and improves their adaptability to various contexts. In the context of a DRP workflow (and ML in general), three key components of modular code typically include data preparation, model training, and inference runs \cite{partin_deep_2023}. These steps can either be integrated into a single script or separated into individual scripts. Proper documentation is crucial for clearly understanding these steps, enabling users to reproduce, adapt, or extend the workflow. This is especially important when working with external data or modifying components of the model.

A comprehensive README file in a code repository should clearly explain each step of the DRP workflow, providing sufficient detail for users to understand how to prepare data, train the model, and run inference. Command line examples and code snippets are invaluable for illustrating how to execute the three steps. The quality of documentation significantly impacts how quickly and easily users can implement the workflow without needing to refer back to the associated manuscript or other sources for guidance.


Below is the scoring system used to evaluate the documentation of code repositories regarding the DRP workflow and its components:

\textbf{Score of 4}.
The README provides a comprehensive explanation of each step in the DRP workflow, including data preparation, model training, and inference runs. The steps may be combined into a single script or separated into different scripts, as long as the README clearly outlines the purpose and execution order of each script. The key is to ensure that the documentation enables users to understand and reproduce the entire workflow seamlessly, regardless of the script organization.
This ensures that users can quickly implement the steps without requiring additional information from the manuscript.

\textbf{Score of 3}.
The README covers the entire workflow, but code examples are only provided for some steps. The explanation is mostly clear, but might have minor gaps in detail. 

\textbf{Score of 2}.
The README provides only a partial description of the workflow, with significant details missing. Code examples are minimal or absent, and users are likely to encounter obstacles in implementing the code without additional guidance.

\textbf{Score of 1}.
The README does not provide any workflow explanation, nor does it offer examples or other guidance on executing the code.

\subsection{Data Availability and Preprocessing} \label{sec:data_and_preproc_methods}

Data preprocessing involves transforming raw data, including 'omics representations, drug features, and drug response values, into model-ready data. Scripts that perform this transformation are essential for assessing reusability in DRP models. Comprehensive preprocessing scripts enable users to test models with new data and compare results from models trained on various datasets. If a repository provides only preprocessed datasets without the corresponding raw data and preprocessing scripts, it becomes challenging, if not impossible, to adapt the model for other data sources. In certain cases, repositories share preprocessed datasets without explaining how the data were created. In this study, we excluded models due to a lack of available raw data or because the preprocessing steps were too vague to understand.

Below is grading criteria we used to assess the quality and completeness of data preprocessing:

\textbf{Score of 4}.
All necessary preprocessing scripts are available in the repository, and the data required to run these scripts is either included in the repository or detailed instructions are provided for download. This ensures that users can recreate the entire data preparation process from raw data to model-ready data.

\textbf{Score of 3}.
All necessary preprocessing scripts are available in the repository, but the data required to run these scripts, or instructions to download it, are missing. While users can understand the preprocessing steps, they cannot fully reproduce the data preparation process without the data.

\textbf{Score of 2}.
The repository contains limited preprocessing scripts, leaving significant gaps in the data preparation process.

\textbf{Score of 1}.
The repository does not provide any preprocessing scripts, making it virtually impossible to use the model with external datasets or recreate the original data preparation process.

\subsection{Reproducing key performance scores} \label{sec:reproduce_methods}

The methods and results reported in research papers should ideally be reproducible by external researchers. Reproducing results helps ensure that a model behaves as expected and can be extended for further analysis, such as comparison with other models. In this section, we do not aim to reproduce all the results reported in the papers. Instead, our goal is to reproduce only key performance metrics from each paper listed in Table \ref{table:list_of_models}, ensuring the proper usage of models.
In addition, to further assess the ease of reusing the models, we recorded the total time spent setting up the environment and getting the model running without errors. 

The reproduced scores are referred to here as \textit{reference scores}. These scores can serve as a reference, ensuring that the model's behavior has not changed after adapting the code for other analyses.
The reference scores used in this analysis include common metrics for assessing the performance of prediction models, such as \textit{R squared}, Pearson correlation coefficient (PCC), Spearman correlation coefficient (SCC), and area under the receiver operating curve (AUROC). Since not all publications presented all scores, the choice of which scores to recreate was subjective.

Our goal was to minimize code modifications to preserve the authentic code structure as much as possible. Whenever feasible, we preserved the original data splits (such as folds or random seeds) and data partitions (like cell line-blind, drug-blind, or random). Changes to these partitions can substantially influence a model's performance, though delving into such effects is outside the scope of this study.

\begin{table}[h]
\caption{A summary of the scoring system for model reproducibility \ref{sec:selection_criteria}.}
\begin{tabular}{l | c p{8cm}}
\toprule
 & Score & Description \\
\midrule
Software Environment (section \ref{sec:sw_env_methods}) & 1 & No information on required software environment.\\
 & 2 & Only a subset of required packages is specified.\\
 & 3 & All packages are listed, but version numbers are limited or missing.\\
 & 4 & All packages with specific version numbers are listed in the requirements file or README.\\
\midrule
Code Modularity (section \ref{sec:modularity_methods}) & 1 & The README has no workflow explanation or code examples.\\
 & 2 & The README offers only a partial workflow description, with limited or no code examples.\\
 & 3 & The README covers the entire workflow, but code examples are provided for only some steps.\\
 & 4 & The README comprehensively explains every step of the DRP workflow, including code examples.\\
\midrule
Data Availability and Preprocessing (sec. \ref{sec:data_and_preproc_methods}) & 1 & No preprocessing scripts are provided.\\
 & 2 & Only some preprocessing scripts are available.\\
 & 3 & All preprocessing scripts are included, but data or download instructions are missing.\\
 & 4 & All preprocessing scripts and data (or clear download instructions) are provided.\\
\bottomrule
\end{tabular}
\label{tab:scoring_system}
\end{table}

\section{Results} \label{sec:results}
\begin{figure}
    \centering
    \includegraphics[width=1\linewidth]{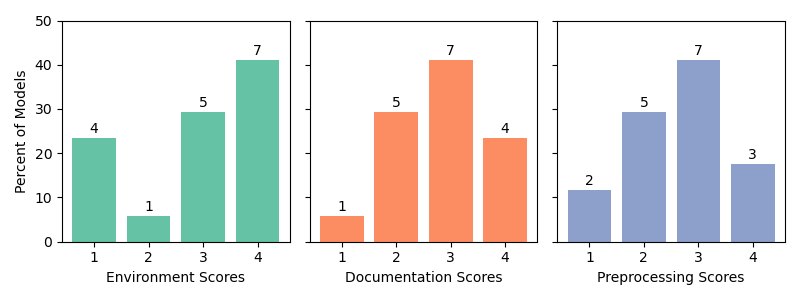}
    \caption{
    Histograms showing the distribution of model scores for each category, with the number of models shown above each bar.
    }
    \label{fig:score_hist}
\end{figure}

The results summarized below are structured according to the reusability scoring system and reproducibility of references scores described in section \ref{sec:methods}. Table \ref{tab:scoring_results} and Figures \ref{fig:score_hist} and \ref{fig:ref_scores} present the statistical details of these results.

\begin{table}[t]
\centering
\resizebox{\columnwidth}{!}{
\begin{tabular}{@{}llcccccl@{}}
\toprule
 & Model & Year & Environment Score & Modularity Score & Preprocessing Score & Time (h) \\
\midrule
1 & BiG-DRP & 2022 & 3 & 4 & 1 & 3 \\
2 & DeepCDR & 2020 & 3 & 4 & 2 & 8 \\
3 & DeepTTA & 2022 & 4 & 2 & 3 & 5 \\
4 & DRPreter & 2022 & 4 & 3 & 3 & 0.5 \\
5 & DrugCell & 2020 & 4 & 4 & 2 & 0.1 \\
6 & DrugGCN & 2021 & 3 & 3 & 3 & 11 \\
7 & DualGCN & 2022 & 4 & 2 & 1 & 12 \\
8 & GraphCDR & 2021 & 4 & 2 & 2 & 2 \\
9 & GraphDRP & 2022 & 2 & 3 & 4 & 8 \\
10 & HiDRA & 2021 & 1 & 3 & 3 & 10 \\
11 & IGTD & 2021 & 4 & 3 & 3 & 0.5 \\
12 & MOLI & 2019 & 1 & 2 & 3 & 8 \\
13 & PathDSP & 2021 & 1 & 2 & 2 & 2 \\
14 & SWNet & 2021 & 4 & 4 & 4 & 3 \\
15 & tCNNs & 2019 & 1 & 1 & 4 & 2 \\
16 & TCRP & 2021 & 3 & 3 & 2 & 10 \\
17 & TGSA & 2022 & 3 & 3 & 3 & 4 \\
\bottomrule
\end{tabular}
}
\caption{\textbf{List of models and scoring results}. Scoring results obtained as part of the reusability and reproducibility analysis are listed for the models. The scoring metrics include software environment (section \ref{sec:sw_env_results}), code modularity (section \ref{sec:doc_results}), and data availability and preprocessing (section \ref{sec:data_and_preproc_results}). The column Time indicates the time required to set up the environment and run the model without errors.}
\label{tab:scoring_results}
\end{table}


\subsection{Software Environment} \label{sec:sw_env_results} 
Seven of 17 models successfully met the full criteria, with the software environment containing all required packages and their respective version numbers (see Table \ref{tab:scoring_results} and Figure \ref{fig:score_hist}).
For models with incomplete or missing environment descriptions, we determined compatible package versions through trial and error or by directly contacting the authors.
Outdated package versions can also increase time required for installation. DRPreter specified a version of PyTorch no longer supported by Conda, so we manually pip installed the package.
According to some users, installing compatible dependencies was the most time-consuming step in the entire study.
Conversely, DrugCell, which was containerized with Docker, required the least amount of time to set up and run.
We observe that rebuilding environments can be a time-consuming and error-prone process, even for experienced users, often requiring trial and error to resolve dependency conflicts and ensure proper configuration.

\subsection{Code Modularity} \label{sec:doc_results}
According to our scoring system described in Table \ref{tab:scoring_system}, the highest score is awarded to models that provided command line examples for each step of the DRP workflow in their documentation, including preprocessing, training, and inference.
These models are more likely to be reusable in various contexts.
As shown in Table \ref{tab:scoring_results} and Figure \ref{fig:score_hist}, only four models (24\% of the total) met this requirement by providing command-line examples for the entire workflow. Another seven models had documentation that explained the complete workflow but lacked command line examples for the individual steps. The top-scoring models, and particularly SWNet, described data files, included installation instructions, and gave code examples for different use cases. It is important to note that a qualitative assessment of the documentation was a factor in the model selection process (see Section \ref{sec:selection_criteria}). Consequently, many models were discarded during the initial selection phase due to inadequate documentation.

\subsection{Data Availability and Preprocessing} \label{sec:data_and_preproc_results} 
Based on our analysis, eight models provided both the data and the preprocessing scripts required to transform it into the form required by the model, earning a score of 4 in this category. 
The repositories varied in the information they provided, ranging from model-ready datasets without preprocessing scripts to full preprocessing steps but without the raw data required for preprocessing. 
Only two models scored 1, lacking both a preprocessing script and raw data to generate model-ready files.
Common challenges in this category included hard-coded paths, missing preprocessing functions, and improperly structured files in the directory tree, all of which generally required troubleshooting.
Table \ref{tab:scoring_system} and Figure \ref{fig:score_hist} show the distribution of scores, reflecting availability of data and preprocessing scripts.

\subsection{Reproducing key performance scores} \label{sec:reproduce_results}
The time taken to re-implement the models, excluding the actual model training time, varied widely, ranging from less than an hour to over twelve hours, as reported in Table \ref{tab:scoring_results}, with a median time of 4 hours.
The reported times for individual models were influenced, in part, by the familiarity of the model curators with the DL framework and model architecture. Furthermore, challenges such as setting up the required environment, missing code or data files, and minor bugs like hard-coded paths were reported as significant factors contributing to the longer setup times. 
In contrast, DrugCell, which was containerized using Docker, proved to be the fastest model to set up.

Due to the difference in evaluation schemes and reported metrics across publications, the reproducibility results between various models are not directly comparable. 
We quantify reproducibility using the percentage deviation of the reproduced score from the original one with 

\begin{equation}
d = 100 * (m_{r} - m_{o}) / m_{o},
\end{equation}


where \(m_{o}\) and \(m_{r}\), are respectively, the original and the reproduced scores, and \(d\) is the percentage deviation.

\begin{figure}
    \centering
    \includegraphics[width=1\linewidth]{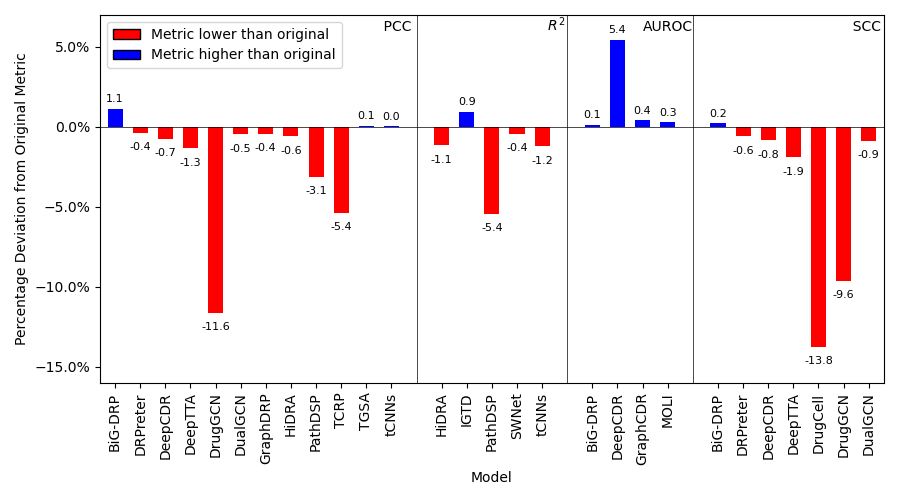}
    \caption{Percentage deviation of reproduced metrics, where the indicated metric is published. A negative score (red) means the reproduced value is lower than the published value. Pearson's correlation coefficient (PCC), coefficient of determination ($R^2$), area under the curve (AUC), and Spearman correlation coefficient (SCC) shown.}
    \label{fig:ref_scores}
\end{figure}

As shown in Figure \ref{fig:ref_scores}, eleven models demonstrated deviations of less than 5\% for all selected metrics. Of the remaining six models, BiG-DRP and DeepTTA exhibited substantial deviation in root mean squared error (RMSE) but not in other metrics. 
For BiG-DRP, the differences in RMSE between the published and reproduced results appear to be due to discrepancies between the model implemented in the code repository and the results published in the paper. Specifically, the paper reports RMSE using z-score normalized IC50 values, whereas the repository produces non-normalized scores. 
DeepTTA showed an approximately 10\% deviation in RMSE, but the deviations in PCC and SCC were relatively small (less than 2\%).
The performance of TCRP was evaluated on a per-drug basis \cite{ma2021tcrp}. To assess reproducibility, we selected the drug with the highest reported PCC value from the paper. Our analysis revealed a deviation of slightly over 5\% between the reported and reproduced PCC values for that specific drug.
Lastly, our analysis of the DrugCell model showed the highest deviation of approximately 14\% when comparing the original and reproduced SCC scores.

\section{Discussion and Recommendations} \label{sec:discussion}

Multiple validation and assessment studies have analyzed various aspects of DRP models, such as generalization across cell-line studies \cite{xia_cross-study_2022}, data scaling properties \cite{partin_learning_2021, branson_comparison_2024}, drug representations \cite{clyde_systematic_2020, baptista_evaluating_2022}, and interpretability \cite{li_interpretable_2023}, while primarily focusing on prediction performance. The current study, however, provides a different view of examining models, specifically geared towards reusability and extensibility rather than prediction performance. The goal here is not to compare models, but rather to understand what aspects of reusability are most overlooked and should receive more attention from developers in the future.
In this paper, we introduce a scoring system to quantify the reusability of prediction models (Table \ref{tab:scoring_system}) and rigorously apply it to 17 DRP models sourced from peer-reviewed literature (Table \ref{table:list_of_models}).
Our analysis reveals that the vast majority of models fall short of perfect scores, indicating potential challenges that external users may encounter when attempting to utilize or build upon these models.

Underspecified software environments force users to determine required package versions through trial and error.
Missing description and examples of the workflow require users to sift through various scripts in the repository in an attempt to piece together a functioning pipeline, further complicating code reusability.
These challenges significantly increase the time needed to understand and reimplement a model.
Notably, data preprocessing steps are frequently omitted, making dataset reconstruction challenging or even impossible, which directly impacts the ability to apply the model to new datasets or extend the model to various analyses.


We find reference scores are a useful tool for diagnosing problems while reimplementing a model, and when altering a model to accept a new dataset or enhance performance. Time to obtain reference scores is significantly reduced by providing a container for the model. 
Most models show moderate underperformance compared to original metrics (Fig. \ref{fig:ref_scores}), but a low reference score does not mean that a model's approach or design is flawed. Rather, the case may be that the publication did not provide sufficient steps to recreate the model without prior knowledge.

%
Challenges reported in this study align with findings from the work of So et al. \cite{so2023reusability}, who investigated the reproducibility and reusability of the TCRP model proposed by Ma et al \cite{ma2021tcrp}. Ma et al. did not provide final hyperparameters, requiring reoptimization and making comparisons between results difficult. Including command line examples with parameters could address this issue. Also, as So et al. noted, some drug results were not replicable. We considered TCRP’s preprocessing code to be incomplete, which is reflected in a score of 2 for the 'Data Availability and Preprocessing' category. 
Our PCC reproducibility score for TCRP, based on KU-55933, is 0.53 compared to the published 0.56 and So et al.'s 0.38. While all results show KU-55933 as the top scorer with similar trends between drugs, the reasons for the substantial differences in PCC are unclear. These reimplementation challenges, though not invalidating the model, make it harder to reuse and limit its overall impact.
  

Despite our efforts to create clear and straightforward guidelines for the scoring system, individual bias may still influence scoring. After the initial submission of scores, we cross-reviewed all entries, identified instances where categories were incorrectly scored, updated the scores, and confirmed them with the original scorers.
As large language models (LLMs) are extensively explored in various scientific domains, they can automate redundant and clearly defined tasks. LLMs can be fine-tuned to auto-score model repositories based on the proposed scoring system, assisting developers in sharing high-quality models and helping the community analyze the potential reusability of public models. 

We observe that many papers introduce novel DRP approaches to improve prediction performance, often claiming state-of-the-art results, while the field lacks agreed-upon benchmark datasets. This suggests that research groups often work in isolation rather than collaboratively, with authors typically not expecting others to adopt and extend their work. To promote rigorous practices and open-source sharing, the field would benefit from clear, standardized guidelines for developing and sharing DRP models. Adhering to these guidelines can mitigate many of the identified issues in this study, enhancing model reusability without imposing significant additional burdens on researchers.
Below, we provide recommendations to achieve this.


\textbf{Recommendation 1}: \textbf{Environment Setup and Containerization}.

The code repository should include clear instructions for recreating the computational environment required for the model. This typically involves providing a requirements.txt file (pip installation) or an environment.yml (Conda environment), ideally accompanied by command-line instructions for environment setup.
Furthermore, including containerization files, such as Dockerfile for Docker or definition file (ending with \textit{.def}) for Singularity, along with instructions for building and running containers can significantly enhance usability and mitigate hardware-related compatibility issues. Specifying the exact versions of key dependencies and Python version is critical.


\textbf{Recommendation 2}: \textbf{Data and Preprocessing Scripts}.

The code repository should include all necessary data preprocessing scripts, and ideally, all the necessary data itself. This includes incorporating feature selection, both by providing explicit feature lists and by automating their selection during preprocessing and dataset generation stages. If data inclusion is restricted due to size or licensing constraints, the repository should provide scripts or clear instructions to facilitate data acquisition. Automating data acquisition and preprocessing eliminates the need for users to manually reproduce complex preprocessing steps based on often incomplete documentation.

\textbf{Recommendation 3}: \textbf{Code Modularity and Documentation}.

To enhance understanding and adaptability, ensure that your code repository is modular, separating key components of the DRP workflow—data preparation, model training, and inference runs—into clearly defined steps.
Whether these steps are integrated into a single script or divided into multiple scripts, the documentation should outline the purpose and execution order of each script.
Include command-line examples to facilitate user implementation and reproduction of the workflow, minimizing the need to refer back to the manuscript or other sources.

\textbf{Recommendation 4}: \textbf{Testing in a Fresh Environment}.

Before publishing your repository, test the complete workflow by cloning it onto a separate system, setting it up, and running the code. This process helps identify smaller issues, such as hard-coded paths or misnamed files, that may not be apparent otherwise. By ensuring the repository functions correctly in a fresh installation, you address key issues and improve the overall robustness of your code.

\section{Conclusion} \label{sec:conclusions}

The lack of proper documentation, modular source code, and full preprocessing information in open-source DRP models hinders their understanding and reuse. This limits their adoption by the scientific community and, consequently, their potential impact on precision oncology.
In this study, we conducted a comprehensive evaluation of 17 DRP models and demonstrate the challenges in their reusability and reproducibility.
Our findings underscore the need for standardized code sharing best practices, not only in DRP but across all AI-enabled scientific models. 
The proposed recommendations aim to foster a more transparent and collaborative research environment, facilitating knowledge sharing and driving progress across various scientific domains.
Ultimately, our recommendations are a starting point for community-developed guidelines; greater engagement from model developers on this topic will lead to a fuller understanding of what makes a model reusable.


\section*{Funding}
The research was part of the IMPROVE project under the NCI-DOE Collaboration Program that has been funded in whole or in part with Federal funds from the National Cancer Institute, National Institutes of Health, Task Order No. 75N91019F00134 and from the Department of Energy under Award Number ACO22002-001-00000.

\section*{Code availability}
This study focuses on the qualitative aspects of reusing and implementing currently available code bases for cancer drug response prediction. The results in this paper do not pertain to the scientific merits, claims, or applicability of the cited research studies. The implementations and modifications made as a result of this effort are available in this GitHub repository: \url{https://github.com/JDACS4C-IMPROVE}.

\bibliographystyle{unsrt}  
\bibliography{references, zotero-alex-partin}


\end{document}